\documentclass[fleqn,usenatbib]{mnras}

\usepackage[T1]{fontenc}

\DeclareRobustCommand{\VAN}[3]{#2}
\let\VANthebibliography\thebibliography
\def\thebibliography{\DeclareRobustCommand{\VAN}[3]{##3}\VANthebibliography}

\usepackage{graphicx}	
\usepackage{amsmath}	
\usepackage{amssymb}	
\usepackage{booktabs}
\usepackage{multicol}
\usepackage{caption}
\usepackage{subcaption}
\usepackage{threeparttable}
\usepackage{tabularx}
\usepackage{array}
\usepackage{tablefootnote}

\newcommand{\hi}{H\,{\sc{i}} } 
\newcommand{\loh}{\log{\! \left(L_{\text{OH}}\right)}}

\newcommand{\sqdeg}{deg$^{2}$ }
\newcommand{\kmps}{\text{km\,s}^{-1}}
\newcommand{\tkmps}{$\text{km\,s}^{-1}$}

\newcommand{\ud}{\mathrm{d}} 

\usepackage{newtxtext,newtxmath}

\title[Efficient selection of lensed OH megamasers]{Efficient selection of gravitationally lensed OH megamasers with MeerKAT and the Square Kilometre Array}

\author[C. B. Button et al.]{
Charissa B. Button$^{1}$\thanks{E-mail: charissa@imago-web.co.za} and 
Roger P. Deane$^{2,1}$
\\
$^{1}$University of Pretoria, Lynwood Road, Hatfield, Pretoria, 0028, South Africa\\
$^{2}$Wits Centre for Astrophysics, University of the Witwatersrand, Jan Smuts Avenue, Johannesburg, 2000, South Africa\\
}

\date{Accepted 2024 January 18. Received 2024 January 17; in original form 2023 November 21}

\pubyear{2023}

\begin{document}
\label{firstpage}
\pagerange{\pageref{firstpage}--\pageref{lastpage}}
\maketitle

\begin{abstract}
There has been a recent resurgence in hydroxyl (OH) megamaser research driven by Square Kilometre Array (SKA) precursor/pathfinder telescopes. This will continue in the lead-up to the SKA mid-frequency array, which will greatly expand our view of OH megamasers and their cosmic evolution over $\gtrsim80$ per cent of the age of the universe. This is expected to yield large scientific returns as OH megamasers trace galaxy mergers, extreme star formation, high molecular gas densities, and potentially binary/dual supermassive black hole systems. In this paper, we predict the distortion to the OH luminosity function that a magnification bias will inflict, and in turn, predict the distortion on the OH megamaser number counts as a function of redshift. We identify spectral flux density thresholds that will enable efficient lensed OH megamaser selection in large spectral line surveys with MeerKAT and SKA. The surface density of lensed galaxies that could be discovered in this way is a strong function of the redshift evolution of the OH megamaser luminosity function, with predictions as high as $\sim$1 lensed OH source per square degree at high redshifts ($z \gtrsim 1$) for anticipated SKA spectral line survey designs. This could enable efficient selection of some of the most highly-obscured galaxies in the universe. This high-redshift selection efficiency, in combination with the large survey speed of the SKA at $\lesssim$1~GHz frequencies and the high magnifications possible with compact OH emission regions ($\mu_{\rm OH} \gg 10$), will enable a transformational view of OH in the universe.
\end{abstract}




\section{Introduction}
\label{sect:Introduction}
Hydroxyl (OH) megamasers are luminous, extragalactic maser sources. As in their Galactic counterparts, the emission in OH megamasers (OHMs) is dominated by the masing lines at 1665 and 1667\,MHz, with much weaker satellite lines at 1612 and 1720\,MHz. However, unlike Galactic OH masers, the emission line at 1667\,MHz is stronger than the emission line at 1665\,MHz, with a typical ratio of the line strengths of 9:5 in local thermodynamic equilibrium \citep[][and references therein]{lo_2005_MegaMasersGalaxies}. Additionally, due to Doppler broadening in massive, often merging galactic systems, OHMs have significantly larger line widths \citep[$\sim$100--1000\,\tkmps;][]{darling_2005_OHMegamasers} than Galactic OH masers. When compared to galaxy scale emission components, OHMs are compact sources, with sizes of order $\sim$100\,pc revealed by high-resolution radio imaging \citep[e.g.][]{pihlstrom_2001_EVNMERLIN, rovilos_2003_ContinuumSpectral, lo_2005_MegaMasersGalaxies}.

Because OHMs require luminous far infrared (IR) radiation to maintain the population inversion that can produce stimulated emission \citep{lockett_2008_Effect53}, they are typically found in the nuclear regions of luminous and ultra-luminous infrared galaxies (LIRGs and ULIRGs), many of which are also major merger systems. Indeed, the integrated OH line luminosity of OHMs is strongly correlated with the far-IR luminosity of the host galaxy, where the correlation follows a super-linear power law,
\begin{equation}
    L_{\text{OH}} \propto \left(L_{\text{FIR}}\right)^{1.2},
    \label{eqn: L_OH L_FIR correlation}
\end{equation}
\citep{baan_1992_IRAS14070, darling_2002_SearchOH, glowacki_2022_LookingDistant, wang_2023_FarInfrared}. However, not all (U)LIRGs host OHMs. The fraction of (U)LIRGs that host OHMs is a strong function of the IR luminosity and increases to about one in three for ULIRGs \citep{lo_2005_MegaMasersGalaxies}. Using HCN and CO observations of a sample of OHMs, \citet{darling_2007_DenseGas} show that high molecular gas densities ($n_{\text{H}_{2}} \gtrsim 10^4\,\text{cm}^{-3}$) are required in addition to strong far-IR radiation for the production of OHMs. Since OHMs reside in LIRGs and ULIRGs, the number density of OHMs should evolve strongly with redshift as the number of (U)LIRGs increases with redshift due to the $\gtrsim 1$~dex increase in the cosmic star formation rate density \citep[e.g.][]{madau_2014_CosmicStar}. Furthermore, the number density of OHMs should also increase with redshift due to the expected increase in the molecular gas fraction and density at high redshift \citep{obreschkow_2009_CosmicDecline}.

OHMs have been demonstrated to be useful tracers for classes of galaxies that are important for understanding several aspects of galaxy evolution. Because they are associated with strong IR radiation and high molecular gas densities, they trace extreme star formation  \citep{darling_2007_DenseGas,lockett_2008_Effect53}. Additionally, since they seem to be produced primarily in major galaxy mergers, they ought to provide an independent probe of the galaxy merger rate \citep{briggs_1998_CosmologicallyDistant}. They are also likely signposts for dual or binary AGN in what are typically obscured environments, especially at higher redshifts.

Some of the first searches for OHMs targeted luminous IR sources with strong radio continuum \citep{baan_1985_FourthOH}. The largest systematic search for OHMs, to date, was conducted with the Arecibo OH megamaser survey \citep{darling_2002_SearchOH}. This survey targeted galaxies selected from the IRAS Point Source Catalogue Redshift Survey \citep[PSCz;][]{saunders_2000_PSCzCatalogue} that were within the declination range $0^{\circ} < \delta < 37^{\circ}$ and that fell within the redshift range $0.1 \leq z \leq 0.23$. Due to the flux limit of the PSCz and the lower redshift limit of Arecibo, the target galaxies were primarily luminous IR galaxies (LIRGS) with $L_{\text{FIR}} \geq 10^{11.4}\,\text{L}_{\odot}$. The Arecibo survey detected 52 new OHMs, almost doubling the number of known OHMs at the time.

Despite this major step forward by the Arecibo OH Survey over two decades ago, the number of known OHMs today is still at a similar level (for more up to date catalogues of OHMs, see \cite{zhang_2014_MiddleInfrared} and \cite{sotnikova_2022_RadioVariability}). Upcoming wide area spectral surveys on MeerKAT and the SKA1-Mid will mark a step-change in the population size. \citet{roberts_2021_OHMegamasers} predicted that the LADUMA survey \citep{blyth_2016_LADUMALooking} alone could detect $\sim 80$ OHMs in its single pointing, while \citet{glowacki_2022_LookingDistant} reported an OHM detection at $z=0.52$ in the LADUMA survey, and \citet{jarvis_2023_Discovery} report the detection of a $z=0.71$ OHM in the MIGHTEE survey, making these the highest redshift detections to date by factors of $\sim2-3$, respectively. It is worth highlighting that the latter, MIGHTEE $z=0.71$ detection, is likely to be strongly lensed, with a magnification factor of $\mu_{\rm OH} \sim 3$. This highlights the opportunities that the SKA1-Mid and its pathfinders, including MeerKAT, will open up for studying both high-redshift and low-luminosity OHMs, which will further our understanding of OHMs and provide useful tracers for understanding aspects of galaxy evolution. 

As the upper redshift limit of cosmic OH is increased by more sensitive instruments, so too is the probability of detecting gravitationally lensed OH megamasers, analogous to what SKA and its precursors/pathfinder will do for the neutral hydrogen line \citep[e.g.][]{deane_2015_StronglyLensed,deane_2016_GravitationallyLensed,blecher_2019_FirstDetection}. A targeted approach to lensed OHMs is described in Manamela et al.~(in prep.), while in this work we describe the untargeted, statistical approach. This relies on the increase in the observed OHM number density due to magnification bias. This distortion to the luminosity function by the sub-population of lensed objects would be most noticeable at high luminosity values because of the exponential decline in the number density at this end, assuming that the luminosity function follows the form of a Schechter function. This increase in the number counts can be exploited to find a peak flux density threshold above which the lensed population dominates over the unlensed population, thereby enabling an efficient approach to a challenging but scientifically rich objective. This selection technique has been used very successfully to discover strong gravitational lenses in far-IR and sub-mm surveys undertaken with \textit{Herschel} and the South Pole Telescope \citep[e.g.][]{negrello_2010_DetectionPopulation, vieira_2013_DustyStarburst, wardlow_2013_HerMESCandidate, negrello_2017_HerschelATLASSample}.

In this paper, we investigate the extent to which a statistical selection approach could be applied to MeerKAT and SKA1-Mid spectral line surveys to identify lensed OHMs. In order to apply the statistical selection approach to OHMs, the OH luminosity function has to be extrapolated to a larger range of luminosities for which it was measured. Section~\ref{sect: MeerKAT and SKA surveys} briefly reviews the relevant surveys that are ongoing with MeerKAT and that are planned for the SKA1-Mid. Section~\ref{sect:OH luminosity function} describes the current constraints on the OH luminosity function and investigates suitable models to extrapolate the function to a larger range in OH luminosity and redshift. Section~\ref{sect:statistical selection approach} reviews the necessary steps for calculating the integrated source counts for OHMs, while Section~\ref{sect:ch4 results and discussion} presents and discusses the results of the lens selection approach. Finally, Section~\ref{sect:ch4 conclusion} presents the conclusions and gives some perspectives on the future outlook for OH studies. Unless otherwise stated, we assume a Planck 2018 cosmological model \citep{planckcollaboration_2020_Planck2018}. 

\section{Overview of MeerKAT and SKA surveys}
\label{sect: MeerKAT and SKA surveys}
Next-generation radio telescopes, such as the Square Kilometre Array (SKA) and its precursors and pathfinders will have dramatically improved instantaneous sensitivity, bandwidth and field of view which will enable the advance of spectral line studies in both \hi and OH. This section describes the surveys that are currently being undertaken with MeerKAT and that have been proposed for the SKA1-Mid. In particular, we detail the expected sensitivity of these surveys.

There are two large survey projects currently being undertaken with MeerKAT that are relevant for spectral line searches for OH emission lines. The first is the MIGHTEE survey \citep{jarvis_2016_MeerKATInternational, maddox_2021_MIGHTEEHIHI} that will survey a sky area of 20\,\sqdeg over four fields at L-band and a smaller region in S-band. Each pointing will have an integration time of $\sim 16$ hrs and the survey should reach a sensitivity of $\lesssim 100$\,$\mu$Jy in a 209~kHz channel. The second is the LADUMA survey \citep{blyth_2016_LADUMALooking} that will spend $\sim 300$\,hrs covering a single pointing on Chandra Deep Field South in L-band and another $\sim 3000$\,hrs on the same field in UHF-band.

A large part of the observing time on SKA1-Mid will be devoted to large survey projects, as has been the case with MeerKAT and ASKAP, the two SKA1-Mid precursors. \citet{staveley-smith_2015_HIScience} outline three prospective tiered surveys, each of a 1000\,hours with survey areas ranging from 400\,\sqdeg to 1\,$\text{deg}^2$. Additionally, \citet{staveley-smith_2015_HIScience} discuss commensal surveys that could still be useful for spectral line science but that could have up to 10\,000 hours of observing time, covering an area of up to of order $\Sigma \sim \pi$\,sr.

In order to calculate the anticipated sensitivity of the SKA1-Mid surveys, we used the estimated $A_{\text{eff}}/T_{\text{sys}}$ values for the SKA1-Mid dishes which are provided by \citet{braun_2019_AnticipatedPerformance}. Additionally, we used the measured mean System Equivalent Flux Density (SEFD) values for a single MeerKAT antenna available on the MeerKAT specifications page to account for the system noise of the MeerKAT dishes at the frequency intervals at which they will contribute to the SKA1-Mid array. From these SEFD values, the sensitivity of the SKA1-Mid for a single pointing can be estimated from the radiometer equation. The estimated sensitivities of the SKA1-Mid surveys, proposed by \citet{staveley-smith_2015_HIScience}, are summarised in Table~\ref{tab: SKA survey sensitivities}. While the survey strategies will only be finalized in the future, these surveys serve as useful, indicative, reference points as we consider searching for lensed OH megamasers in the SKA1-Mid and MeerKAT surveys.

\begin{table*}
    \small
    \centering
    \begin{tabular}{cccccccc}
        \toprule
         Survey   & Area & Time & $\nu$ & FOV & Pointings & Int. time & $\sigma_{\text{S}_{\nu}}$  \\
                & \sqdeg & hrs & MHz & \sqdeg & & hrs & $\mu$Jy/beam \\
        \midrule
        Medium wide & 400 &  1 000 & $950 - 1420$ & 0.94 & 428 & 2.34 & 15.6 \\
        Medium deep & 20 & 1 000 & $950 - 1420$ & 0.94 & 22 & 45.5 & 3.54 \\
        Deep & 1 & 1 000 & $600 - 1050$ & 1.71 & 1 & 1 000 & 1.30 \\
        \midrule
        All sky & 20 000 & 10 000 & $950 - 1420$ & 0.94 & 21 357 & 0.47 & 34.9 \\
        Wide & 5 000 & 10 000 & $950 - 1420$ & 0.94 & 5 340 & 1.87 & 17.5 \\
        Ultra deep & 1 & 10 000 & $450 - 1050$ & 1.71 & 1 & 10 000 & 0.45 \\
        \bottomrule
    \end{tabular}
    \caption[Flux density sensitivity of SKA1-Mid surveys]{Flux density sensitivities of the SKA1-Mid surveys proposed in \citet{staveley-smith_2015_HIScience}. The field of view is calculated at the maximum frequency quoted for the survey. Column 7 gives the effective integration time per pointing calculated from the number of pointings and the total number of hours for the survey. The flux density sensitivity is calculated for a single pointing at the central frequency of the survey and for a rest-frame velocity width of 200\,$\kmps$ which ranges between 950 and 470\,kHz for \hi between redshifts of 0 and 1.}
    \label{tab: SKA survey sensitivities}
\end{table*}

\section{The OH megamaser luminosity function}
\label{sect:OH luminosity function}
The OH luminosity function, $\Theta(L_{\text{OH}})$, is the comoving number density of OHMs as a function of their OH luminosity. However, it is often more convenient to express the luminosity function as the comoving number density per logarithmic luminosity interval rather than linear luminosity interval. In this case, the luminosity function is denoted as $\Phi(L_{\text{OH}}$). The OH luminosity function can be calculated either directly from OHM number counts, or indirectly from the FIR luminosity function, assuming a $L_{\text{OH}}$--$L_{\text{FIR}}$ correlation and an OHM abundance in FIR-luminous galaxies. Both approaches were used in the early work on the OH luminosity function; \citet{baan_1991_ActiveNuclei} used the OHM number counts that were available at the time to derive an OH luminosity function, while \citet{briggs_1998_CosmologicallyDistant} inferred the OH luminosity function indirectly from the FIR luminosity function and the $L_{\text{OH}}$--$L_{\text{FIR}}$ correlation. \citet{darling_2002_OHMegamaser} used the OHM detections in the Arecibo OH Megamaser Survey to estimate the OH luminosity function directly. Although the survey detected OHMs over the luminosity range $1.4 < \loh < 4.2$, \citet{darling_2002_OHMegamaser} used an error-weighted least squares approach to fit a simple power law to the data points with luminosities in the range $2.2 \leq \loh \leq 3.8$. Using this approach, they find the following OH luminosity function (assuming a cosmological model where $H_0 = 75\,\text{km\,s}^{-1}\,\text{Mpc}^{-1}$, $\Omega_{\text{M}} = 0.3$, and $\Omega_{\Lambda} = 0.7$),
\begin{equation}
    \Phi_{\text{a}}(L_{\text{OH}}) = \left(9.8_{-7.5}^{+31.9} \times 10^{-6} \right) \left(L_{\text{OH}}\right)^{-0.64\pm0.21} \text{\, Mpc}^{-3} \text{\,dex}^{-1}.
    \label{eqn: OH LF D&G2002}
\end{equation}
Later, \citet{roberts_2021_OHMegamasers} used a Markov chain Monte Carlo approach in a re-analysis of the OH luminosity function using a power law model and the same data as \citet{darling_2002_OHMegamaser}. They obtained the following parameter estimates (scaled to the same cosmological model as used in \citet{darling_2002_OHMegamaser}),
\begin{equation}
    \Phi_{\text{b}}(L_{\text{OH}}) = \left(2.58 \pm 0.47 \times 10^{-6}\right) \left(L_{\text{OH}}\right)^{-0.50\pm0.13} \text{\, Mpc}^{-3} \text{\,dex}^{-1}.
    \label{eqn: OH LF R+2021}
\end{equation}

\begin{figure}
    \centering
    \includegraphics[width=1.\columnwidth]{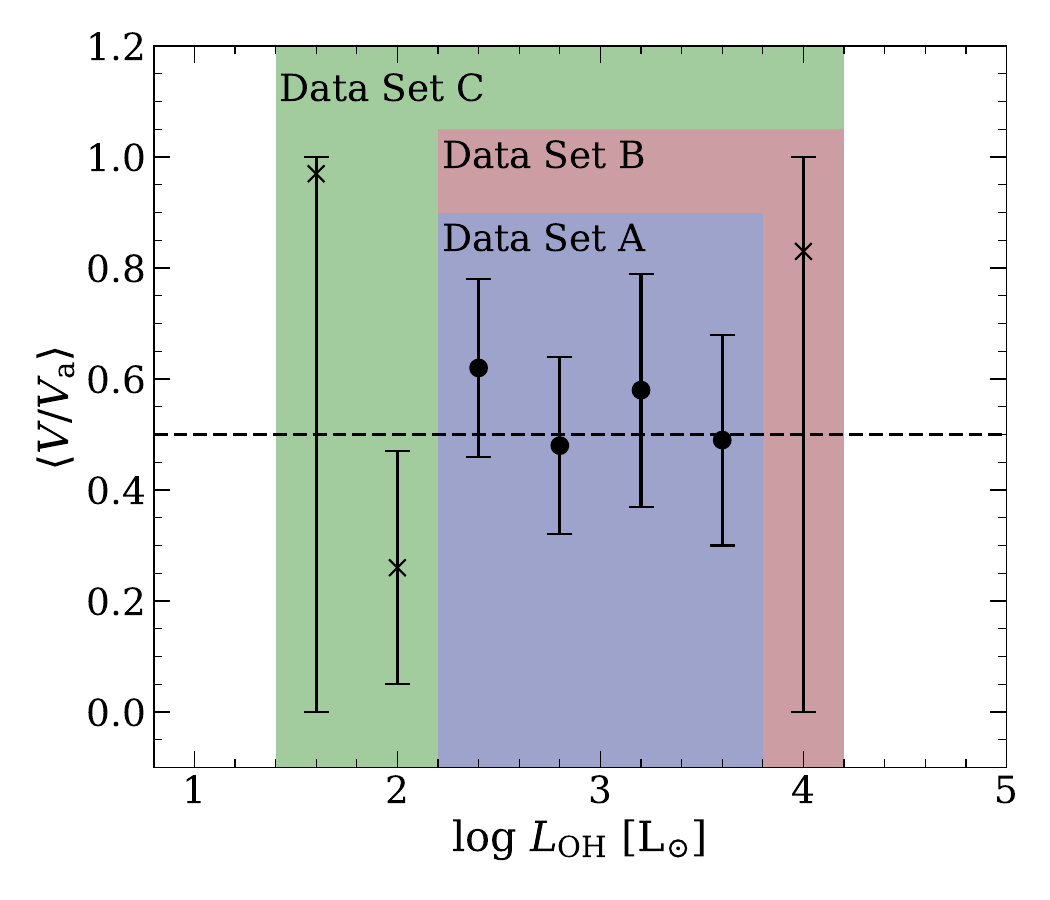}
    \caption[Data sets for modelling the OH luminosity function]{Average $V$ to $V_{\text{a}}$ ratios for each luminosity bin. The $\langle V/V_{\text{a}} \rangle$ ratio is used to test for uniform sampling; a value of 0.5 indicates that the volume available within a luminosity bin is well-sampled. The coloured intervals indicate the different data sets that are used in the model selection tests in Section~\ref{sect:refitting OH LF}, where the Data Set A shows the data points that were used in the analyses performed by \citet{darling_2002_OHMegamaser} and \citet{roberts_2021_OHMegamasers}. Figure reproduced from the data points reported in \citet{darling_2002_OHMegamaser}.}
    \label{fig:ave V_Va ratios}
\end{figure}

As mentioned, in their analyses both \citet{darling_2002_OHMegamaser} and \citet{roberts_2021_OHMegamasers} only used data points within the luminosity range $2.2 \leq \loh \leq 3.8$. The data outside of this range were excluded as they argued that the OHM detections in these luminosity bins did not uniformly sample the available cosmological volumes. To test for uniformity, \citet{darling_2002_OHMegamaser} used the $\langle V/V_{\text{a}}\rangle$ test from \citet{schmidt_1968_SpaceDistribution} where a uniformly distributed sample has a $\langle V/V_{\text{a}}\rangle$ value of 0.5. Here, $V$ is the survey volume at the redshift of each detected OHM and $V_{\text{a}}$ (also referred to as $V_{\text{max}}$ by other authors) is the maximum available volume for each OHM given the survey sensitivity and the luminosity of the OHM. Figure~\ref{fig:ave V_Va ratios} shows the $\langle V/V_{\text{a}}\rangle$ values and uncertainties estimated by \citet{darling_2002_OHMegamaser}. The points marked by crosses indicate the luminosity bins centred on $\loh = 1.6$, $2.0$, and $4.0$ that were excluded in the analyses by \citet{darling_2002_OHMegamaser} and \citet{roberts_2021_OHMegamasers}. The $\langle V/V_{\text{a}}\rangle$ values for the luminosity bins centred on $\loh = 1.6$, $2.0$ and $4.0$ are consistent with a value of 0.5 within 1.2$\sigma$; however, each of the bins centred on $\loh = 1.6$, and $4.0$ had only one detection which means that the uncertainties on the $\langle V/V_{\text{a}}\rangle$ values for these bins are on order of 1.

With well-sampled number counts, luminosity or mass functions of galaxy properties are typically well described by a Schechter function \citep{schechter_1976_AnalyticExpression}. While the OH luminosity function and its evolution are poorly constrained at present, upcoming observations with SKA precursors/pathfinders will significantly increase the number of known OH megamasers across a wider range of luminosity and redshift.

In what follows, we perform Bayesian model selection between a power law and Schechter function model. Taking a Bayesian approach, our analysis includes all the detections from the Arecibo OHM survey. However, we also perform this same model selection with subsets of the data to compare with previous work and the effect of excluding data. We also present a typical parametrization of the redshift evolution of the OH luminosity function.

\subsection{A Bayesian approach to modelling the OH luminosity function}
\label{sect:refitting OH LF}
In this section, we model the OH luminosity function with both a power law and a Schechter function using three combinations of the data presented in \citet{darling_2002_OHMegamaser}. We used \textsc{PyMultiNest} \citep{buchner_2014_XraySpectral} to constrain the model parameters for both a power law model and a Schechter model, as well as to compute the Bayesian evidence for each model required for model selection. The 2-parameter power law model is given by
\begin{equation}
    \Phi(L_{\text{OH}}) = a \log{\left(\frac{L_{\text{OH}}}{\text{L}_{\odot}} \right)} + b,
\end{equation}
while the 3-parameter Schechter model is given by
\begin{equation}
\begin{split}
    \Phi(L_{\text{OH}}) \ud \loh = & \phi^* \ln(10) \left( \frac{L_{\text{OH}}} {L_{\text{OH}}^*} \right)^{\alpha + 1} \\
                                   & \exp{\left(- \frac{L_{\text{OH}}} {L_{\text{OH}}^*} \right)} \ud \loh.
\end{split}
\label{eqn: OH LF schechter model}
\end{equation}
Since \citet{darling_2002_OHMegamaser} did not include all data points, we carried out three investigations using different subsets of data points (labelled Data Sets A, B, and C, see Fig.~\ref{fig:ave V_Va ratios}). First, we modelled the luminosity function using the same data points as used by \citet{darling_2002_OHMegamaser}, i.e. the luminosity bins centred on $\loh = 2.4$, $2.8$, $3.2$ and $3.8$. This is referred to as Data Set~A in Table~\ref{tab:OH LF refitting results} and Figure~\ref{fig:ave V_Va ratios}. Second, in Data Set~B we extend the data subset to include the high luminosity point at $\loh = 4.0$ and repeat the same modelling approach. Third, Data Set~C includes all the luminosity bins and we again perform the same modelling with this data set. Figure~\ref{fig:ave V_Va ratios} shows the different subsets of data points used in the three investigations.

The results of the parameter estimation and model selection are presented in Table~\ref{tab:OH LF refitting results}. It shows, for each of the three investigations, the Bayes factor comparing the evidence for the Schechter model to the evidence for a power law model, as well as the median values of the posterior probability distributions for each parameter in the two models. In the first two investigations that exclude some data points, there is only weak or inconclusive support for a power law model over a Schechter function, using the Jeffrey's scale interpretation \citep[see e.g.][]{trotta_2017_BayesianMethods}. When considering all the data points, the Schechter model is preferred, although the Bayes factor shows inconclusive support for the Schechter model in this case. 

\begin{figure*}
    \centering
    \includegraphics[width=0.65\textwidth]{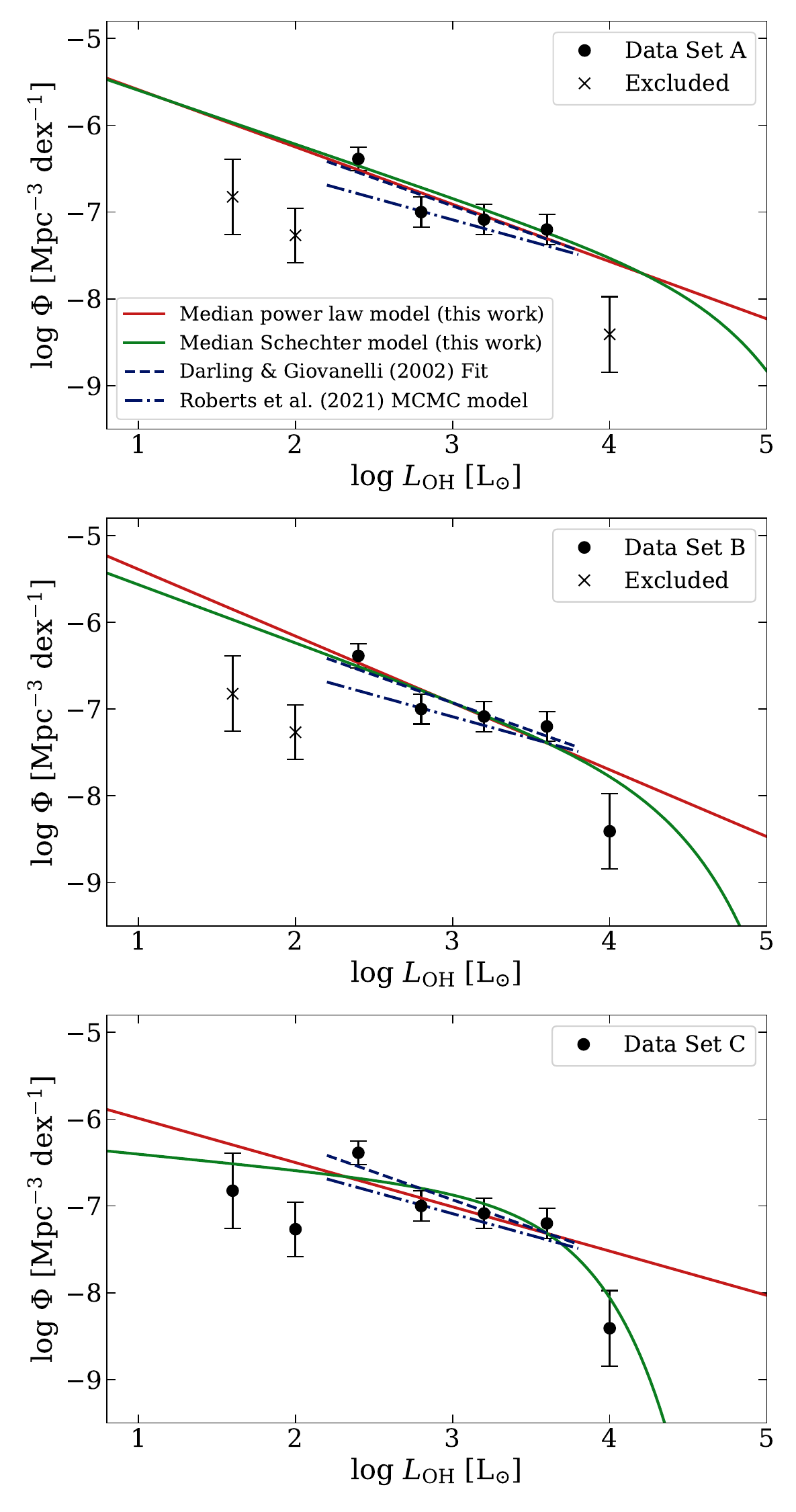}
    \caption[Results from the parameter estimation of the OH luminosity function]{Results from the parameter estimation of the OH luminosity function. The top panel shows the results for the parameter estimation using Data Set~A, the middle panel shows the results for the parameter estimation using Data Set~B, and the bottom panel shows the results for the parameter estimation using Data Set C. In the top panels, the points that are excluded from the parameter estimation are marked with crosses. In this figure, the same cosmological parameters are assumed as in \citet{darling_2002_OHMegamaser} and the parameters from \citet{roberts_2021_OHMegamasers} are scaled accordingly.}
    \label{fig:Refitted OH LF}
\end{figure*}

The median posterior models for the three investigations are shown in Figure~\ref{fig:Refitted OH LF}. It is interesting to note that for Data Set A, the nested sampling approach produces a similar power law to that found by \citet{darling_2002_OHMegamaser}. Similarly, for Data Set C, the nested sampling approach produces a power law that is consistent to within the uncertainties with the power law produced by the MCMC approach used by \citet{roberts_2021_OHMegamasers}. The fact that the power law model from Data Set C agrees with the model found by \citet{roberts_2021_OHMegamasers} can be viewed as argument against the exclusion of the data points. The differences in the derived power law parameters for Data Set~A are expected to be due to differences in the MCMC algorithm used in \citet{roberts_2021_OHMegamasers} and the nested sampling algorithm used in this work. The latter has been shown to be a more robust, albeit computationally expensive approach \citep{skilling_2004_NestedSampling}. Given that nested sampling computes the Bayesian evidence while MCMC does not, we would expect superior results with our approach.

Overall, the results show that the model selection between a power law model and a Schechter model is inconclusive given the current data. There is weak or inconclusive support that a power law model is preferred over the narrower selected luminosity range used in \citet{darling_2002_OHMegamaser}. However, it is unclear that this model can be extrapolated to either lower or higher luminosities, with the latter being a key point of interest in this paper. 

\begin{table*}
    \centering
    \begin{tabular}{c c c c c c}
        \toprule
         Data Set & Luminosity range & $\Delta \log {L_{\text{OH}}}$   & $\ln B$ & Schechter  & Power law   \\
          & $\log{ \left( L_{\text{OH}}/\text{L}_{\odot} \right) }$ & & & & \\
         \midrule
         A                    & 2.2--3.8            & 0.4 & -1.3    & $\log{\phi^*} = -8.29$     & $a = -0.66$ \\
                              &                     &     &         & $\log{\! \left(L_{\text{OH}}^*\right)} = 4.76$  & $b = -4.93$ \\
                              &                     &     &         & $\alpha = -1.62$           &             \\
         \midrule
         B                    & 2.2--4.2            & 0.4 & -0.7  & $\log{\phi^*} = -8.16$       & $a = -0.77$ \\
                              &                     &     &       & $\log{\! \left(L_{\text{OH}}^*\right)} = 4.33$  & $b = -4.62$ \\
                              &                     &     &       & $\alpha = -1.67$                                &             \\
         \midrule
         C                    & 1.4--4.2            & 0.4 & 0.6   & $\log{\phi^*} = -7.23$                          & $a = -0.51$ \\
                              &                     &     &       & $\log{\! \left(L_{\text{OH}}^*\right)} = 3.59$  & $b = -5.48$ \\
                              &                     &     &       & $\alpha = -1.18$                                &             \\
         \bottomrule
    \end{tabular}
    \caption[Summary of results from the OH luminosity function parameter estimation]{Summary of results from the OH luminosity function parameter estimation. The first column indicates the luminosity range of the data points used in the parameter estimation. The second column lists the width of the luminosity bins. The third column shows the natural logarithm of the Bayes factor calculated so that a positive value indicates support for the Schechter model. The fourth and fifth column show the median values of the posterior parameter distributions for the Schechter and power law model respectively. Note that these results assume the same values for the cosmological parameters as in \citet{darling_2002_OHMegamaser}.}
    \label{tab:OH LF refitting results}
\end{table*}

In the remainder of this work, the OH luminosity function is taken to be modelled by a Schechter function, with parameter estimation based on all the available data in all luminosity bins. There are several reasons that motivate this choice. First, within a Bayesian framework, all data points should be included even where there are large uncertainties. Second, when all the data points are considered, a Schechter model is marginally favoured even though the Bayes factor is inconclusive. Third, a wide range of luminosity functions of all kinds are generally well modelled by a Schechter function and it is reasonable to expect that when the OH luminosity function is measured over a larger range of luminosities, it will be no different. Fourth, if the $L_{\text{OH}}$--$L_{\text{FIR}}$ correlation holds to higher luminosities, we expect that the OH luminosity function will follow the far-IR luminosity function which is well modelled by a Schechter function. 

We note that the lens selection approach we utilise in this paper is sensitive to the exponential steepening of the luminosity function at high luminosities, which is poorly constrained at this point. However, this will become more tightly constrained as more high redshift OHMs are detected with MeerKAT and the SKA1-Mid. In anticipation thereof, we consider what the possibilities are for detecting both lensed and unlensed OHMs en route to a new era in OHM studies.

The Schechter parameters for the OH luminosity function that are used in this work, converted to the Planck 2018 cosmology \citep{planckcollaboration_2020_Planck2018} are $\log \phi^* = -7.36 \pm 0.31$, $\log{\left(L_{\text{OH}}^*\right)} = 3.68 \pm 0.52$ and $\alpha = -1.18 \pm 0.26$.

\subsection{Parametrization of the redshift evolution}
\label{sect:OHLF z evolution}
As discussed, the number density of OHMs is expected to evolve strongly with redshift for several reasons \citep[e.g.][]{briggs_1998_CosmologicallyDistant, darling_2002_OHMegamaser}. First, since OHMs have been observed to be associated with luminous far-IR radiation, the redshift evolution of the OH luminosity function is expected to be strongly influenced by the evolution of the number density of (U)LIRGs. Although (U)LIRGs in the local universe are predominantly major merger systems, at high redshift the (U)LIRG population also contains normal star forming galaxies \citep[e.g.][]{daddi_2010_VeryHigh}. Thus, the number density of (U)LIRGs should evolve with redshift due to both the increasing star formation activity \citep[e.g.][]{madau_2014_CosmicStar} and the evolution of the merger rate. However, the merger rate of galaxies is still highly uncertain \citep[see e.g.][]{mundy_2017_ConsistentMeasure}. Some studies find that for massive galaxies, the merger rate increases \citep[e.g.][]{bluck_2012_StructuresTotal}, while others find that the merger rate is constant or even decreases \citep[e.g.][]{williams_2011_DiminishingImportance, newman_2012_CanMinor} with selection biases seen to be a major hurdle to reconcile. Second, since OHMs are also associated with dense molecular gas regions, the redshift evolution of the OH luminosity function should also be influenced by the increased mid-plane pressure and dust temperatures of molecular gas in galaxies at higher redshifts \citep{darling_2007_DenseGas,lockett_2008_Effect53}. The molecular gas is expected to have a higher density at higher redshifts due to the galaxy size evolution \citep{gunn_1972_InfallMatter} and due to the predicted increase in the $\text{H}_2/\text{\hi}$ mass ratio with redshift \citep{obreschkow_2009_CosmicDecline}.

Although the redshift evolution of the number density of (U)LIRGs and the increasing density of molecular gas should strongly influence the redshift evolution of the OH luminosity function, at this point it is difficult to quantify this evolution of the OH luminosity function. Therefore, in this work we parametrize the redshift evolution by including a factor of $(1+z)^{\beta_{\text{OH}}}$, as is commonly used to parametrize the redshift evolution of a variety of luminosity or mass functions \citep[e.g. the \hi mass function,][]{pan_2020_MeasuringMass}. The OH luminosity function is then given by
\begin{equation}
    \Phi(L_{\text{OH}}, z) = (1+z)^{\beta_{\text{OH}}} \phi^* \ln(10) \left( \frac{L_{\text{OH}}} {L_{\text{OH}}^*} \right)^{\alpha + 1} \exp{\left(- \frac{L_{\text{OH}}} {L_{\text{OH}}^*} \right)},
\end{equation}
where the exponent, $\beta_{\text{OH}}$, of the $(1+z)$ term is referred to as the evolution parameter in this work. We assume this evolution parameter incorporates a number of relevant, introduced effects, including the contribution from the merger rate, the increase in the abundance of (U)LIRGs, and the increase in the molecular gas density mid-plane pressure. Because this evolution is largely unknown at this stage, we investigate values of $\beta_{\text{OH}} = 0$, $2$, $4$ and $6$ in order to cover a wide range of evolution scenarios. This is similar to the approach taken by \citet{darling_2002_OHMegamaser} to model what they isolated as the merger rate. These different redshift evolution scenarios will be constrained as observations from MeerKAT and SKA1-Mid become available and increase the number of known OHMs to sample sizes with sufficient statistical power, however, for this paper they provide the necessary functional form to explore the lensed OH number counts at high redshift. 

\section{Statistical lens selection approach}
\label{sect:statistical selection approach}
\subsection{Lensing probability}
\label{sect:lensing probability}
The effect of gravitational lensing on the observed luminosity function depends on the probability that a background source at a given redshift is lensed and by a given magnification factor. Here, we follow the method outlined in \citet{perrotta_2002_GravitationalLensing} in order to calculate this probability, which depends on the mass profile of the foreground lens, on the mass and redshift distribution of the foreground lenses, the redshift of the background source, and the cosmological parameters which we fix to the values measured by \citet{planckcollaboration_2020_Planck2018}. Following \citet{perrotta_2002_GravitationalLensing} and \citet{wardlow_2013_HerMESCandidate}, we assume that the foreground lenses are dark matter haloes. \citet{wardlow_2013_HerMESCandidate} found that their predictions did not have the necessary precision to discriminate between the singular isothermal sphere (SIS) and Navarro-Frenk-White (NFW) density profiles. Additionally, the lensing quantities of an SIS density profile are simpler than the lensing quantities of an NFW density profile. For these reasons, we adopt SIS profiles here. For an individual lens, the mass surface density will be enhanced near its nucleus by a baryonic component, however, following other authors, we focus on the dark matter lensing potential for this statistical, rather than individual lens analysis.

The probability that a background source is lensed depends on the lens cross-section which is the area in the source plane within which the magnification is larger than a given value, $\mu_{\text{min}}$, and is given by the equation \citep{lima_2010_LensingMagnification},
\begin{equation}
    \Omega(\mu, z_{\text{L}}, z_{\text{S}}, M_{\text{vir}}) = \int_{\mu > \mu_{\text{min}}} \ud \beta^2,
    \label{eqn:Lens cross-section definition}
\end{equation}
where $\Omega$ is the lens cross-section, $z_{\text{L}}$ is the redshift of the foreground lens, $z_{\text{S}}$ is the redshift of the background source, $M_{\text{vir}}$ is the virial mass of the foreground lens and $\beta$ is the source position. In the case of an SIS profile the cross-section becomes
\begin{equation}
    \Omega(\mu, z_{\text{L}}, z_{\text{S}}, M_{\text{vir}}) = \pi \beta^2(\mu).
    \label{eqn:Lens cross-section SIS}
\end{equation}

Given the lens cross-section, the probability that a background source is lensed with a magnification factor greater than $\mu$ is given by the fraction of the area of the source sphere where the magnification is greater than $\mu$ \citep{perrotta_2002_GravitationalLensing}:
\begin{equation}
\begin{split}
    P(\mu, z_{\text{S}}) = \frac{1}{4 \pi D^{2}_{\text{A}}(z_{\text{S}})} & \int_0^{z_{\text{S}}} \ud z_{\text{L}} \frac{\ud V}{\ud z_{\text{L}}} 
                            \\ & \int \ud M_{\text{vir}} \, \Omega(\mu, z_{\text{L}}, z_{\text{S}}, M_{\text{vir}}) \frac{\ud N}{\ud M_{\text{vir}} \, \ud z_{\text{L}}}.
    \label{eqn: prob source is lensed}
\end{split}
\end{equation}
Here $D_{\text{A}}(z_{\text{S}})$ is the angular diameter distance at the redshift of the source, $\ud V/ \ud z$ is the comoving volume element per unit redshift and solid angle, $\Omega(\mu, z_{\text{L}}, z_{\text{S}}, M_{\text{vir}})$ is the lens cross-section, and $\ud N/\ud M_{\text{vir}} \, \ud z_{\text{L}}$ is the comoving number density of the lenses, which, following \citet{perrotta_2002_GravitationalLensing} is given by the Sheth and Tormen dark matter halo mass function \citep{sheth_2001_EllipsoidalCollapse} and is implemented in the \textsc{Halomod}\footnote{https://halomod.readthedocs.io} package \citep{murray_2013_HMFcalcOnline, murray_2021_TheHaloModOnline}. This expression for the probability is only valid in the limit of non-overlapping cross-sections where $P \ll 1$. However, strong lensing has a very small probability since it requires the source and the lens to be very closely aligned ($\theta \lesssim 1$\,arcsec) and it is rare that a source is lensed by more than one foreground mass. That said, massive galaxies are typically clustered; however, we do not take this into consideration since systematic modelling uncertainties (e.g. OHM source size) far outweigh this second-order effect.

The magnification probability distribution is given by the differential probability,
\begin{equation}
    p(\mu, z_{\text{S}}) = -\frac{\ud P(\mu, z_{\text{S}})}{\ud \mu},
    \label{eqn: prob distribution}
\end{equation}
where $P(\mu, z_{\text{S}})$ is given by Equation~\ref{eqn: prob source is lensed}. Interestingly, $p(\mu, z_{\text{S}}) \propto \mu^{-3}$ at high magnifications irrespective of the lens model.

\subsection{Integrated counts of OH sources}
\label{sect:Integrated counts of OH sources}
The effect of lensing on the observed luminosity function can then be calculated from the magnification probability distribution. Since the magnification is equal to the ratio of the observed flux density to the actual flux density, the observed or apparent luminosity is
\begin{equation}
    L_{\text{OH},\,\text{app}} = \mu L_{\text{OH}}.
\end{equation}
So, if all the sources were lensed by a factor of $\mu$, the observed number density will be related to the intrinsic number density by \citep{pei_1995_MagnificationQuasars}
\begin{equation}
    \Theta'(L_{\text{OH}}) = \frac{1}{\mu} \Theta \left( \frac{L_{\text{OH}}}{\mu} \right),
    \label{eqn: distorted linear OH LF at single mu}
\end{equation}
where $\Theta (L_{\text{OH}})$ is the number density calculated for intervals of $L_{\text{OH}}$ and the factor of $1/\mu$ takes into account the amount by which the magnification changes the width of the luminosity bins. However, it is often more convenient to calculate the number density on intervals of $\log{L_{\text{OH}}}$, as is done in Equation~\ref{eqn: OH LF schechter model}. In this case, the observed number density, $\Phi'(L_{\text{OH}})$ is related to the intrinsic number density by
\begin{equation}
    \Phi'(L_{\text{OH}}) = \Phi \left( \frac{L_{\text{OH}}}{\mu} \right),
    \label{eqn: distorted OH LF at single mu}
\end{equation}
since the magnification factor does not affect the width of the logarithmic luminosity intervals.

When the sources are lensed by a range of magnification factors with a probability distribution given by Equation~\ref{eqn: prob distribution}, then the observed number density is given by
\begin{equation}
    \Phi'(L_{\text{OH}}, z_{\text{S}}) = \int_{\mu_{\text{min}}}^{\mu_{\text{max}}} \ud \mu p(\mu, z_{\text{S}}) \Phi \left( \frac{L_{\text{OH}}}{\mu}, z_{\text{S}} \right).
    \label{eqn: OH LF modified by mag bias}
\end{equation}
Here, the value of $\mu_{\text{min}}$ is restricted to the strong lensing regime ($\mu_{\text{min}}>2$) by the assumption in Equation~\ref{eqn: prob source is lensed} that the lens cross-sections do not overlap, while the value of $\mu_{\text{max}}$ is predominantly limited by the maximal solid angle of the source.

The integrated source counts of OH megamasers are given by
\begin{equation}
    N(>S_{\nu, \, \text{peak}}) = \int_{z_{1}}^{z_{2}} \ud z \int_{L_{\text{min}}}^{{\infty}} \ud \log\!{L_{\text{OH}}} \, \Phi(L_{\text{OH}}, z) \frac{\ud V_c}{\ud z},
    \label{eqn: OH Integrated source counts}
\end{equation}
where $L_{\text{min}}$ is the integrated luminosity of the OH spectral line corresponding to a given peak flux density and $\ud V/ \ud z$ is the differential comoving volume. In the above equation, if the integrated source counts are calculated for the unlensed population, $\Phi(L_{\text{OH}}, z)$ is the OH luminosity function. On the other hand, if the source counts are calculated for the lensed population, $\Phi'(L_{\text{OH}}, z)$ is the OH luminosity function modified by the magnification bias, as given in Equation~\ref{eqn: OH LF modified by mag bias}. Thus, in order to calculate the integrated source counts, we need to specify the maximum magnification, the integrated luminosity corresponding to a given peak flux density, and the redshift interval.

\begin{figure*}
    \centering
    \includegraphics[width=1.0\textwidth]{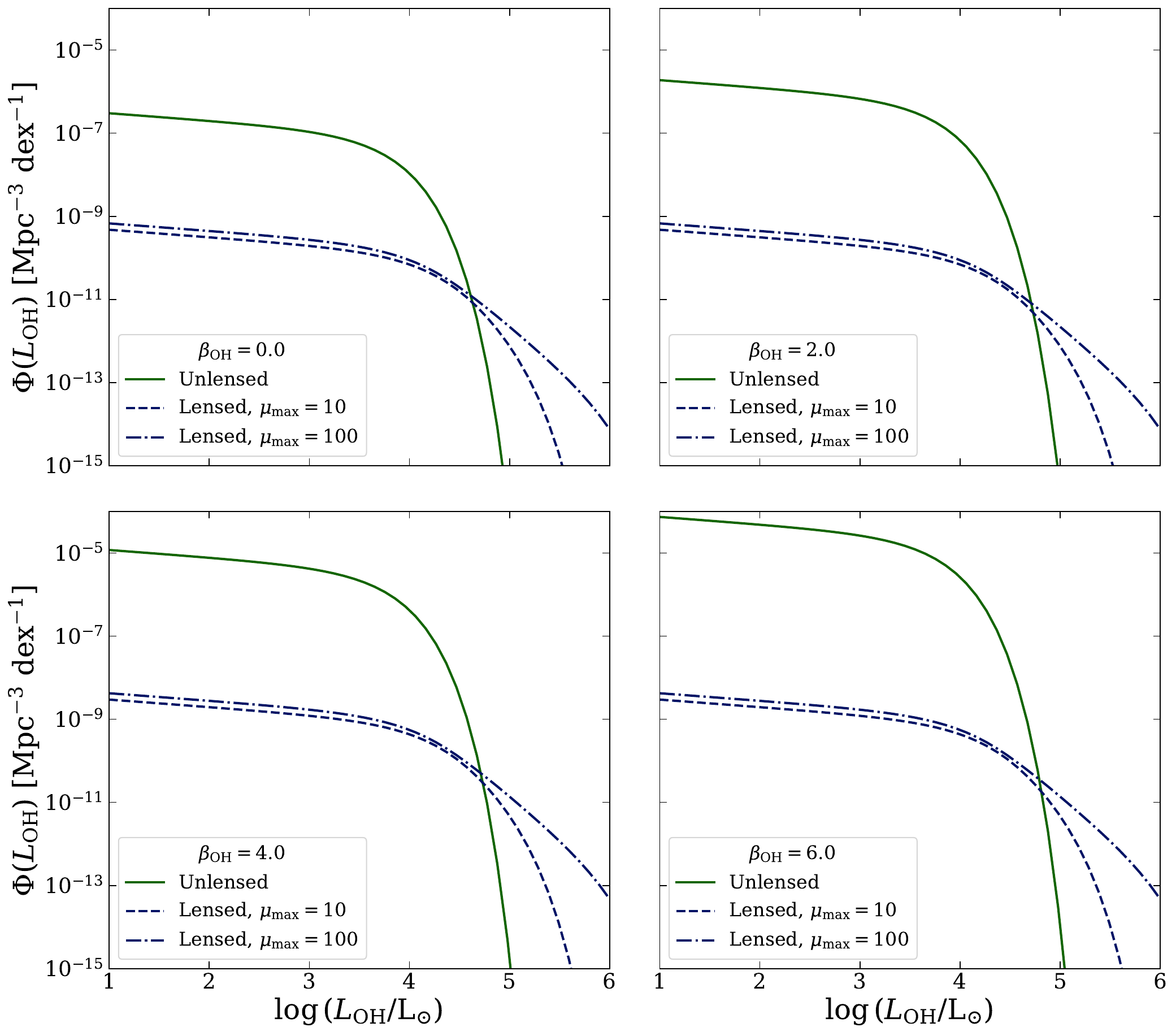}
    \caption[Lensed OH luminosity function]{Comparison of the OHM luminosity function modified by the magnification bias with the unlensed luminosity function for four values of the evolution parameter. The lensed sources are assumed to be at a redshift of 1.5.}
    \label{fig: Lensed vs unlensed OHM LF}
\end{figure*}

As discussed, OHMs are typically  $\lesssim$100\,pc in extent \citep{rovilos_2003_ContinuumSpectral,lo_2005_MegaMasersGalaxies}, which is considerably smaller than most emission components in their host galaxies. This smaller spatial extent should result in much higher magnification factors compared to the magnification factors considered in, for example the \hi case, particularly for galaxy-scale Einstein radii \citep[e.g.][]{deane_2015_StronglyLensed, blecher_2019_FirstDetection}. In order to take into account this expectation that OHMs can have high magnifications ($\mu \gg 10$) and to be consistent with the modelled magnifications seen in similarly sized emission components in the literature \citep[e.g. \textsc{H\,ii} regions,][]{kneib_2011_ClusterLenses}, maximum magnification factors in the range $\mu_{\rm max} = 10$--$100$ are considered here. Figure~\ref{fig: Lensed vs unlensed OHM LF} shows the effect of the magnification bias on the OH luminosity function for different values of the redshift evolution parameter and for maximum magnification factors in the range 10--100. These plots indicate that it is only at high luminosities that the OH luminosity function distorted by the magnification bias for a maximum magnification of 100 differs significantly from the OH luminosity function with a maximum magnification of 10. At the point where the distorted OH luminosity function intersects the unlensed OH luminosity function, the results for the two values of the maximum magnification are similar. A similar result is seen in the integrated source counts that are plotted in Figure~\ref{fig:Integrated source counts OHMs}. This could change as the OH luminosity function constraints improve, however, they show that our predictions are relatively insensitive to this assumption with current models.

The integrated luminosity of the OH spectral line is related to the peak flux density through the integrated flux and the definition of the luminosity distance. Assuming that the spectral line has a rectangular profile, the integrated flux is related to the peak flux density as
\begin{equation}
    S_{\text{int}} = \frac{\nu_{\text{OH}}}{c(1+z)} \Delta V_{\text{rest}} S_{\nu}^{\text{peak}},
    \label{eqn: OH integrated flux ito peak flux density}
\end{equation}
where $\nu_{\text{OH}}$ is the rest-frequency of the OH spectral line, $z$ is the redshift of the OH megamaser and $\Delta V_{\text{rest}}$ is the rest-frame velocity width. Constraints on the velocity width of higher redshift OHMs do not exist, but for consistency with previous predictions by \citet{darling_2002_SearchOH} a constant line width of $150\,\kmps$ is assumed in this paper, however, note that much broader profiles are seen in the low redshift universe. Then, from the definition of the luminosity distance, the integrated flux is related to the integrated luminosity by 
\begin{equation}
    L_{\text{int}} = 4\pi S_{\text{int}} D_{L}^2(z).
    \label{eqn: OH luminosity ito integrated flux}
\end{equation}
Combining Equations \ref{eqn: OH integrated flux ito peak flux density} and \ref{eqn: OH luminosity ito integrated flux}, the luminosity can be expressed in terms of the peak flux density as
\begin{equation}
    L_{\text{int}} = \frac{4 \pi \nu_{\text{OH}}}{c(1+z)} \Delta V_{\text{rest}} S_{\nu}^{\text{peak}} D_{\text{L}}^2.
    \label{eqn: OH luminosity ito peak flux density}
\end{equation}

The SKA1-Mid Band~1 receivers will extend down to a frequency of 350\,MHz, so in principle, OHMs could be detected out to a redshift of $z \sim 3.7$. Hence, the integrated number counts in this work are considered for redshifts up to 3.7. This redshift range is divided into smaller redshift intervals and the number counts are calculated on each interval. It is useful to consider smaller redshift intervals in order to investigate how the number counts change with redshift, even though the number counts are smaller, since a smaller redshift interval corresponds to a smaller volume (at a given central redshift). Here, we consider redshift intervals of width $\Delta z = 0.10$ and $0.37$. This also accounts for practical data processing considerations, as well as the fact that RFI is often clustered in specific windows.

\section{Results and discussion}
\label{sect:ch4 results and discussion}
\begin{figure*}
    \centering
    \includegraphics[width=\textwidth]{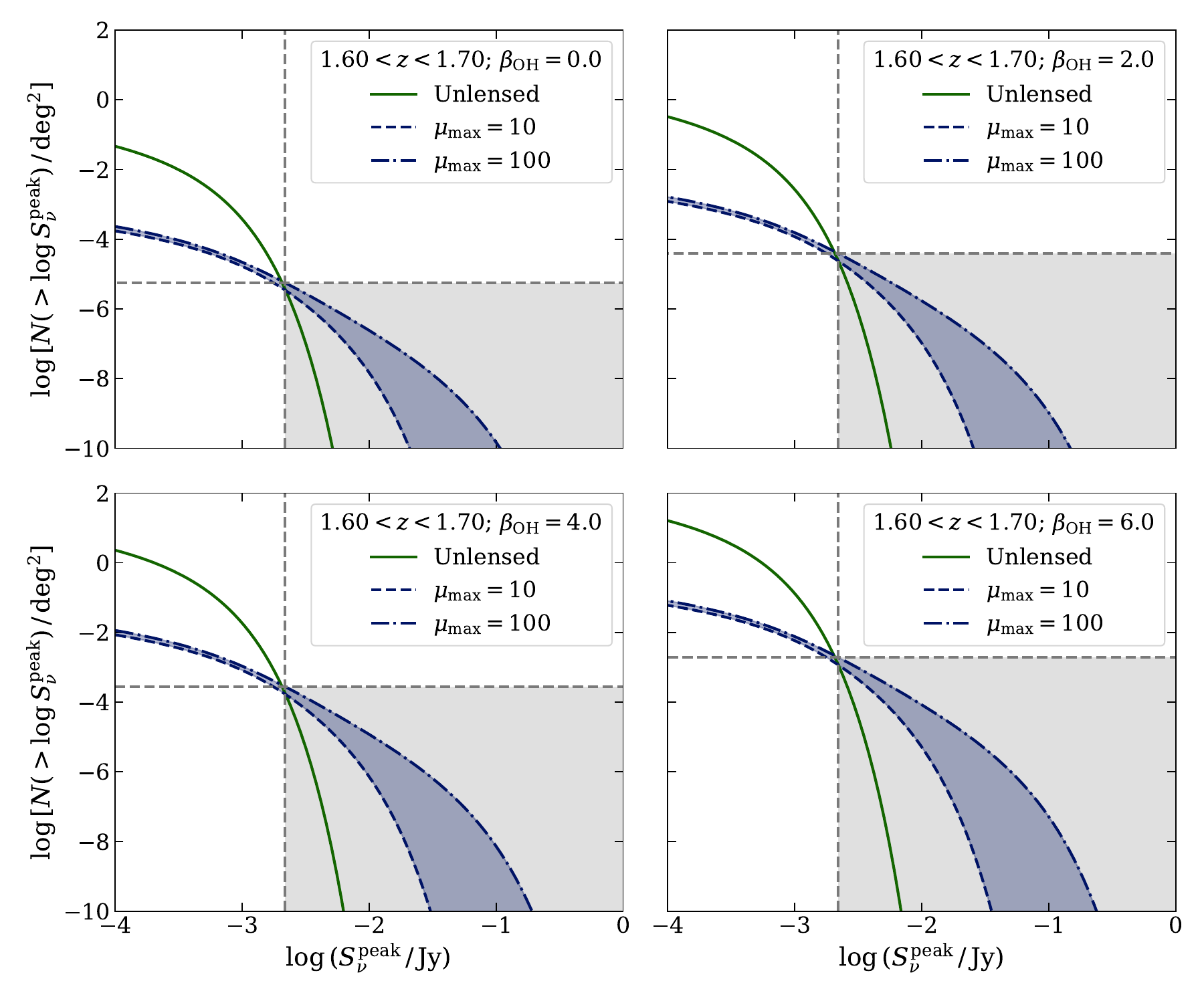}
    \caption[Lensed and unlensed integrated OHM source counts]{Integrated source counts for the lensed (shown in blue) and unlensed (shown in green) OH megamaser populations integrated over the redshift interval, $1.5 \leq z \leq 1.6$. The lensed source counts are calculated at two values of the maximum magnification, $\mu_{\text{max}} = 10$ and $\mu_{\text{max}} = 100$. The grey shading shows the region where the lensed number counts are equal to or exceed the unlensed number counts, making this the region of efficient lens selection, even without multi-wavelength information to exclude containment non-lensed galaxies. The points marked by the intersection of the vertical and horizontal grey dashed lines are referred to as the source count equality points.}
    \label{fig:Integrated source counts OHMs}
\end{figure*}

This section presents the results of applying this selection approach to OHMs. As discussed in the previous sections, the integrated counts of the lensed and unlensed populations are calculated for redshift intervals of width $\Delta z = 0.1$ and $0.37$ within the range $0 \leq z \leq 3.7$ and for values of the evolution parameter of $\beta_{\text{OH}} = 0$, $2$, $4$, and $6$. This results in 47 redshift intervals for each value of the evolution parameter. We do this in order to provide a practical sense of how these searches can be carried out, while also accounting for our ignorance on $\beta_{\text{OH}}$ and the OH luminosity function. In Figure~\ref{fig:Integrated source counts OHMs}, we show the lensed and unlensed integrated counts for each value of $\beta_{\text{OH}}$ in an example redshift bin of $1.6 < z < 1.7$, which is accessible by observed frame frequencies that lie just inside the MeerKAT UHF band ($\nu_{\rm obs} \sim 630$~MHz). Here, the grey shading indicates the region where the lensed number counts are equal to or exceed the unlensed number counts. For the sake of clarity, the point of intersection between the lensed integrated source counts (for $\mu_{\text{max}} = 100$) and the unlensed integrated source counts (indicated by the intersection of the grey dashed lines) are referred to as the \emph{source count equality points}. A sample selected at the flux density threshold indicated by the vertical dashed line would contain lensed and unlensed sources in a ratio of 1:1.

The OHM surface density is a strong function of the evolution parameter and is expected to increase by orders of magnitude in even the more conservative estimates. OHM searches with MeerKAT and the SKA1-Mid should help to constrain the observed OHM surface density as a function of redshift, which will, in turn, enable the redshift evolution of the OHM number density to be constrained.

\begin{figure*}
    \centering
    \includegraphics[width=1\textwidth]{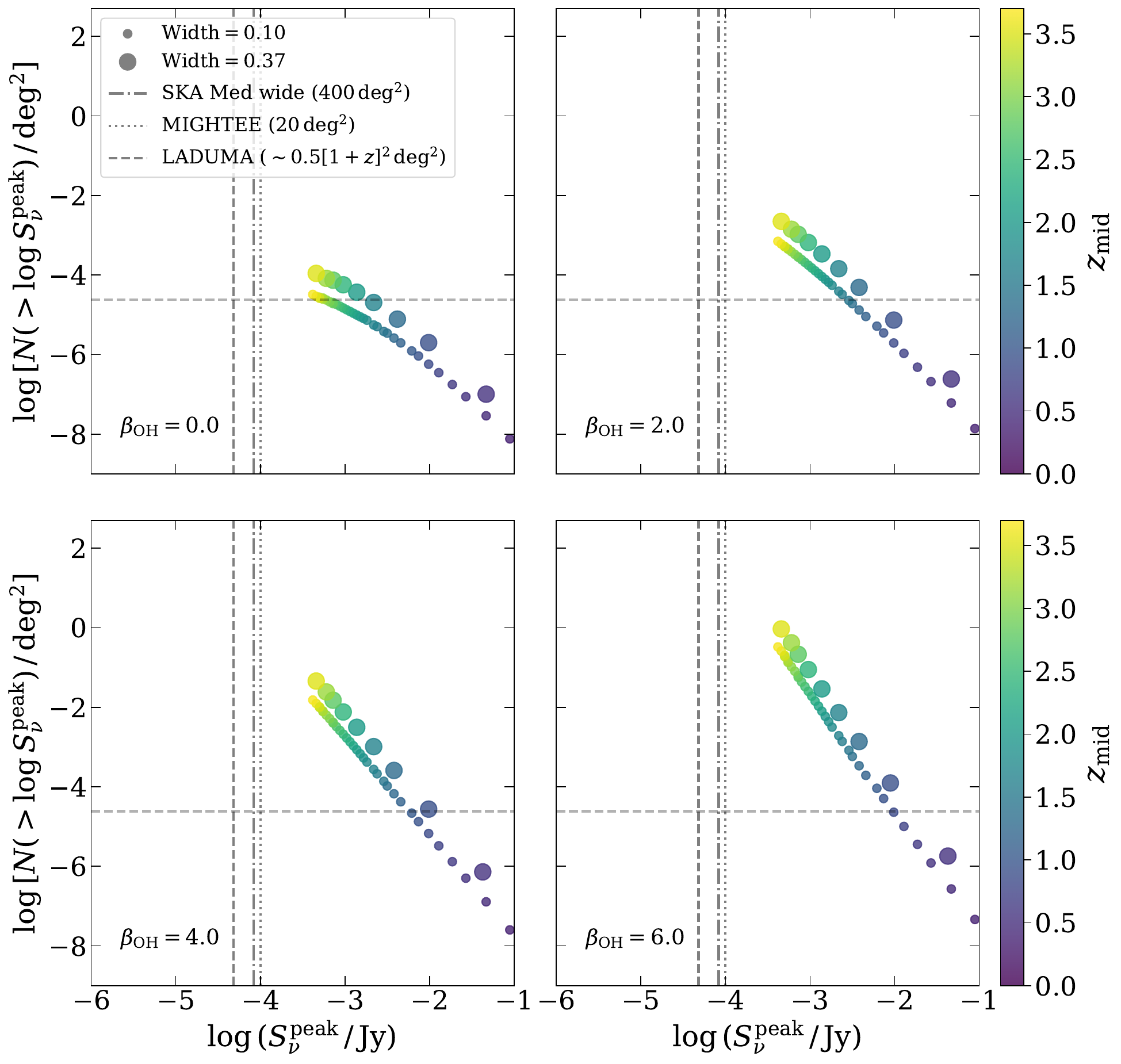}
    \caption[Source count equality points]{The source count equality points for all the redshift intervals considered, at the different values of the evolution parameter, $\beta_{\text{OH}}$, as indicated on each panel. The centre of the redshift interval is indicated by the colour and the size of the points indicates the width of the redshift interval. The approximate sensitivities (assuming a rest-frame velocity width of 150\,\tkmps) for the MIGHTEE and LADUMA surveys of MeerKAT, as well as for the SKA1-Mid Medium Wide survey are shown by the vertical lines. The horizontal dashed line indicates the limiting source surface density of 1 source in the whole sky area.
    }
    \label{fig:Cross-over points OHMs with surveys}
\end{figure*}

The extent to which this method could be useful for selecting lensed OHMs in the SKA1-Mid surveys can be explored by plotting the source count equality points as a function of the central redshift and the width of the redshift interval for the different values of $\beta_{\text{OH}}$. These source count equality points are shown in Figure~\ref{fig:Cross-over points OHMs with surveys} where the vertical lines indicate the approximate sensitivities of the SKA1-Mid medium wide survey, the MIGHTEE survey and the LADUMA survey (5-$\sigma \simeq 80\,\mu\text{Jy beam}^{-1}, 100\,\mu\text{Jy beam}^{-1}, 50\,\mu\text{Jy beam}^{-1}$, respectively, assuming a rest-frame velocity width of $150\,\text{km\,s}^{-1}$ \citet{staveley-smith_2015_HIScience, maddox_2021_MIGHTEEHIHI, blyth_2016_LADUMALooking, roberts_2021_OHMegamasers}). The horizontal dashed lines indicate the limit of 1 source in the whole sky.

These source count equality points occur at flux densities that are easily accessible to the SKA1-Mid medium wide survey and the MIGHTEE survey, which has a similar depth to the SKA1-Mid medium wide survey. Since the LADUMA survey only covers an area of $\sim 3\,\text{deg}^2$ at 580\,MHz, its area does not make it an optimal survey to discover lensed OHMs, while the wider areas of the SKA1-Mid medium wide survey and the MIGHTEE survey make them potentially more interesting surveys to consider here. The surface densities of the lensed sources at these intersection points depend strongly on the evolution parameter, but are small ($< 10^{-4}\,\text{deg}^{-1}$) at redshifts of $z\lesssim1$ for all the evolution scenarios. In the case where there is no redshift evolution (i.e. $\beta_{\text{OH}}=0$), the surface density of lensed OHMs at the source count equality points is close to or below the limit of one lensed source in the whole sky for all redshifts. It only starts to exceed this limit at high redshifts ($z \gtrsim 2.5$) and when integrating over the wider redshift interval ($\Delta z = 0.37$). However, for $\beta_{\text{OH}} = 6$, the surface density increases to 1 lensed source per 1\,\sqdeg at high redshifts. This indicates that if the redshift evolution of the OH luminosity function is relatively strong, $\beta_{\text{OH}} \geq 4$, this could be a promising method for selecting lensed OHMs, with lensed OHMs being detectable out to redshifts of $z \sim 3.5$ at a relatively high surface density. On the other hand, if the OH luminosity function does not evolve strongly with redshift, this selection method becomes limited by the low surface density of the lensed OHMs; however, non-detections of any OH sources, whether lensed or unlensed, will provide joint constraints on the evolution parameter as well as the high end of the OH luminosity function. The above is, of course, for the most pessimistic scenario, where no multi-wavelength information is taken into account. Doing so will likely make a significant enhancement to this selection technique and is the subject of future work. In the current paper, we simply explore indicative levels of contaminant removal.

\subsection{Lensed OHM surface densities at higher contaminant ratios}
\begin{figure*}
    \centering
    \includegraphics[width=1\textwidth]{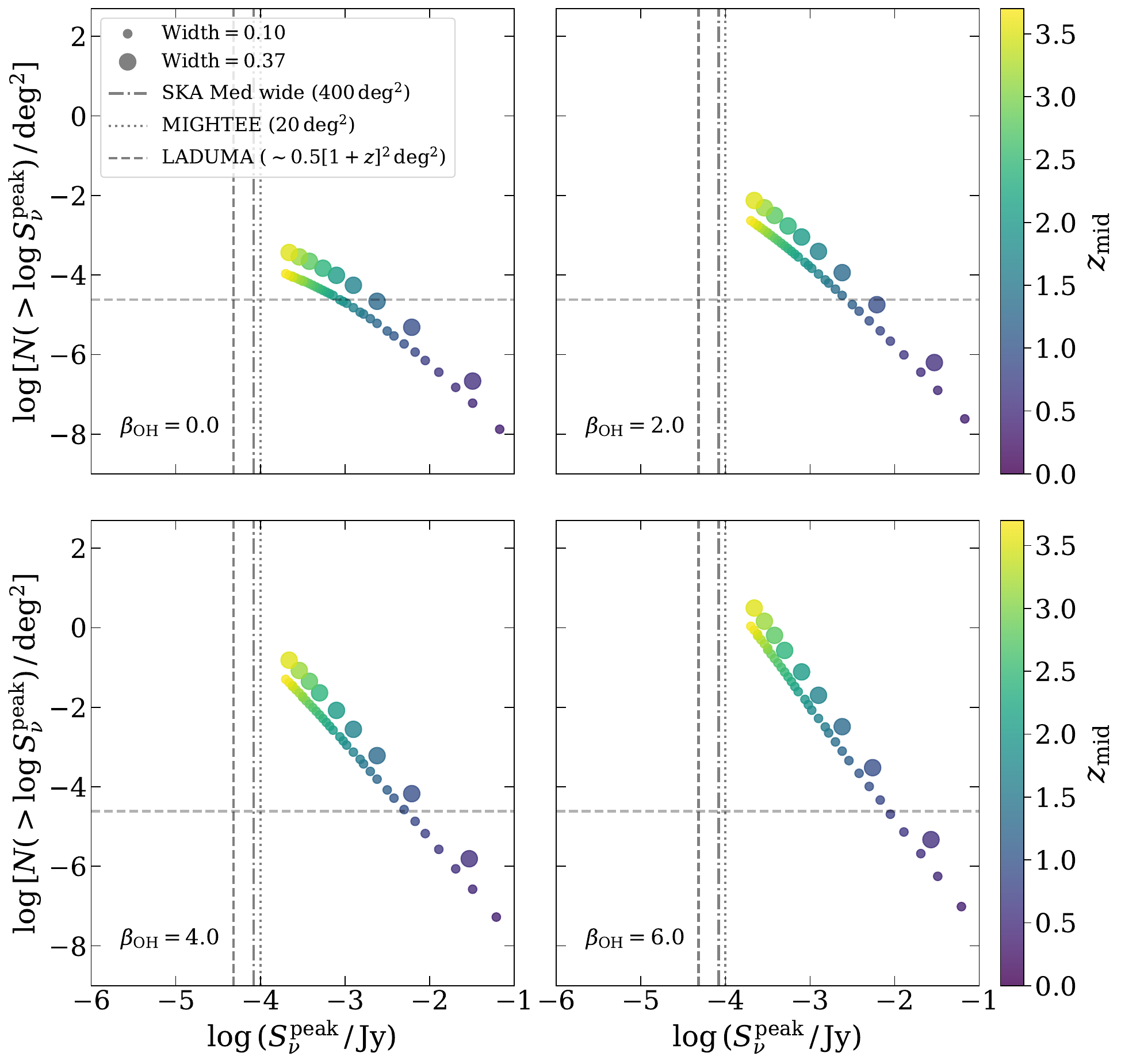}
    \caption[Intersection points for a lensed to unlensed ratio of 1 to 10]{Same as Figure~\ref{fig:Cross-over points OHMs with surveys} but assuming that contaminants at a ratio of 10:1 can be removed.}
    \label{fig:Cross-over points OHMs contaminant removal case 1:10}
\end{figure*}

Multi-wavelength information, such as that from large optical and IR surveys, will greatly assist in identifying lensed OHM candidates in a given sample. If the multi-wavelength selection is efficient, it should be possible to select lensed OHMs from samples that contain lensed to unlensed OHMs in a ratio much less than 1:1, greatly enhancing yield with this technique. We, therefore, investigate how many lensed sources per square degree are expected at lower, indicative, ratios. Figures~\ref{fig:Cross-over points OHMs contaminant removal case 1:10} show the surface densities of the lensed OHMs at flux densities where the unlensed source count is 10 times the lensed source count. We use this value as an indicative exploration of what is possible. These results suggest that for a scenario with no evolution, which is highly unexpected, even a contaminant removal accuracy of 1 in 10 will result in a very low sky density (less than 1 lensed source per 1000\,$\text{deg}^2$) and, hence, have low scientific yield. For stronger evolution parameters ($\beta_{\text{OH}} = 2$, $4$ or $6$), being able to select 1 lensed source out of 10 unlensed sources would naturally increase the surface density of the lensed OHMs. For example, for $\beta_{\text{OH}}=6$ the surface density increases to $\sim 3$ lensed sources per 1\,$\text{deg}^2$, at the highest redshifts. By way of comparison, the lens selection in the \textit{Herschel} ATLAS survey obtained a lens surface density of 0.13 lens candidates per 1\,\sqdeg \citep{negrello_2017_HerschelATLASSample}. This implies that the surface density of lensed OHMs could be significantly larger than the \textit{Herschel} far-IR selection for galaxies in a very similar redshift window centred on cosmic noon ($1 \lesssim z \lesssim 3$). Additionally, selecting lensed OHMs has a significant advantage of immediate spectroscopic confirmation, making follow up observations far more efficient.

\section{Conclusion and future outlook}
\label{sect:ch4 conclusion}
Currently, the number of known OHMs is relatively small (on order of 100) and limited to low redshifts. However, as tracers of extreme star formation and major mergers, these objects will provide useful perspectives on galaxy evolution processes, especially at higher redshifts, as well as providing signposts for dual/binary AGN in obscured environments. The SKA1-Mid and its precursors, especially MeerKAT, will increase the number of OHM detections and advance our understanding of these objects through multi-faceted, multi-wavelength studies of the resultant OHM samples out to significantly larger cosmological distances. In anticipation of these developments, this paper investigates the possibility of selecting lensed OHMs in the ongoing/proposed MeerKAT and SKA1-Mid wide area spectral line surveys.

In the first instance, the OH luminosity function is constrained to both higher and lower luminosities than is carried out in \citet{darling_2002_OHMegamaser} and \citet{roberts_2021_OHMegamasers}. Although the number of current OHM detections are few, using a Bayesian framework with a phenomenological expectation that a Schechter function is an appropriate model for sufficient number counts, we model the OH luminosity function using measurements from \citet{darling_2002_OHMegamaser} with a Schechter function. This chosen model is supported by Bayesian model selection; however, this approach and the resultant luminosity function parameter constraints will be tested as more OHM detections become available in the near future.

Following our Bayesian parameter estimation and model selection, we present the results of the lens selection as applied to OHMs. The prospect of detecting lensed OHMs in wide area spectral line surveys is a strong function of the evolution parameter, $\beta_{\text{OH}}$, which we define to include a wide range of contributing factors, including the major merger rate, the evolution of the IR source counts and the increase in the density of molecular gas with redshift, amongst others. For no or weak evolution, which is highly unlikely, this lens selection method is not effective as the surface density of the lensed OHMs is very small ($< 10^{-4}\,\text{deg}^{-2}$), a scenario easily ruled out by a small sample of detections from a targeted lensed OHM survey (Manamela et al., in prep.). For strong $\beta_{\rm OH}$ evolution, this selection method becomes promising even without any other information as the lensed OHMs should reach a surface density of 1 in 1 \sqdeg at the highest redshifts surveyed with SKA1-Mid Band 1. This surface density can be improved up to $\sim 3$ lensed OHMs per 1\,\sqdeg if ancillary multi-wavelength information can be used to remove contaminants. Clearly, there is great potential scientific yield, which motivates using sophisticated techniques, including machine learning methods, to remove unlensed contaminants, a subject of current research, with a relevant link reported in \citet{roberts_2021_OHMegamasers}, where OHM are seen as the contaminants in separating {\sc H\,i} and OHM sources in the LADUMA survey. These results illustrate how the different redshift evolution scenarios of the OH luminosity function can be tested in the near future as more OHMs are discovered in upcoming wide-area spectral line surveys. We predict that our view of the OH in the universe is about to be transformed beyond previous predictions by the SKA and its precursors/pathfinders, through the power of strong gravitational lensing.

\section*{Acknowledgements}
We thank the anonymous referee for their useful comments. We are grateful to Julie Wardlow, Matt Jarvis, Danail Obreschkow, and Mario Santos for very useful comments and discussions. CBB and RPD acknowledge funding from the South African Radio Astronomy Observatory (SARAO), which is a facility of the National Research Foundation (NRF), an agency of the Department of Science and Innovation (DSI). RPD acknowledges funding by the South African Research Chairs Initiative of the DSI/NRF (Grant ID: 77948). We acknowledge the use of the ilifu cloud computing facility - www.ilifu.ac.za, a partnership between the University of Cape Town, the University of the Western Cape, Stellenbosch University, Sol Plaatje University, the Cape Peninsula University of Technology and the South African Radio Astronomy Observatory. The ilifu facility is supported by contributions from the Inter-University Institute for Data Intensive Astronomy (IDIA - a partnership between the University of Cape Town, the University of Pretoria and the University of the Western Cape), the Computational Biology division at UCT and the Data Intensive Research Initiative of South Africa (DIRISA).

\section*{Data Availability}
No new data were generated or analysed in support of this research.



\bibliographystyle{mnras}
\bibliography{References} 








\bsp	
\label{lastpage}
\end{document}